\documentclass[11pt,a4paper]{article}
\usepackage[utf8]{inputenc}
\usepackage[T1]{fontenc}
\usepackage{geometry}
\usepackage{array}
\usepackage{multicol}
\usepackage{comment}
\usepackage{amsopn,amsmath,amssymb,amsthm,amstext,amsfonts,algorithmic,psfrag}
\usepackage{amssymb}
\usepackage{enumitem}
\usepackage{mathtools}
\usepackage[english]{babel}
\usepackage{hyperref}
\hypersetup{colorlinks=true, citecolor=blue, linkcolor=blue}
\usepackage[round]{natbib}
\setcitestyle{aysep={,}} 
\usepackage{authblk}
\usepackage{graphicx}
\usepackage{mwe}    % loads »blindtext« and »graphicx«
\usepackage{subfigure} % multiple subplots under a plot
\usepackage{float}
\usepackage{caption}%,subcaption}
\usepackage{algorithm}
\usepackage{algorithmic,latexsym}
\usepackage{placeins,enumitem}
\renewcommand{\thetable}{\arabic{table}}

\usepackage{eso-pic}
\newcommand\AtPageUpperMyright[1]{\AtPageUpperLeft{%
 \put(\LenToUnit{0.5\paperwidth},\LenToUnit{-1cm}){%
     \parbox{0.5\textwidth}{\raggedleft\fontsize{9}{11}\selectfont #1}}%
 }}%
\newcommand{\conf}[1]{%
\AddToShipoutPictureBG*{%
\AtPageUpperMyright{#1}
}
}

\title{\LARGE \bf
Cyclone Preparedness, Rescue Operations and Damage Assessment using UAVs
}
\author[1]{Rudrashis Majumder}
\author[2]{Shuvrangshu Jana}
\author[3]{Prathyush P. Menon}
\author[4]{Debasish Ghose}
\author[5]{N. M. Prusty}
\author[6]{Bipasha Mukherjee}
\author[7]{Aditi Ghosh}

\affil[1]{Ph.D. Student, Department of Aerospace Engineering, Indian Institute of Science, Bangalore, 
		{\tt\small rudrashism@iisc.ac.in}}

\affil[2]{Post-doctoral Fellow, Department of Aerospace Engineering, Indian Institute of Science, Bangalore,
		{\tt\small shuvrangshuj@iisc.ac.in}}

\affil[3]{Associate Professor, College of Engineering, Mathematics and Physical Sciences, University of Exeter, UK, 
		{\tt\small p.m.prathyush@exeter.ac.uk}}

\affil[4]{Professor, Department of Aerospace Engineering, Indian Institute of Science, Bangalore, 
		{\tt\small dghose@iisc.ac.in}}		
		
\affil[5, 6, 7]{Humanitarian Aid International, India, 
		{\tt\small $^5$nmprusty51@gmail.com, $^6$bipasha@hai-india.org, $^7$adghosh@gmail.com}}	
		
\date{}                     %% if you don't need date to appear
\begin{document}
\maketitle

\conf{$5^{th}$ World Congress on  Disaster Management\\ IIT Delhi, New Delhi, India, 24-27 November 2021}
%\thispagestyle{empty}
%\pagestyle{empty}

%%%%%%%%%%%%%%%%%%%%%%%%%%%%%%%%%%%%%%%%%%%%%%%%%%%%%%%%%%%%%%%%%%%%%%%%%%%%%%%%

\begin{abstract}
UAV's capability to access remote and inaccessible areas within a quick time can be utilized for effective cyclone management. This paper presents the possible application of UAVs at different stages of cyclone mitigation. The overall system architecture necessary for preparedness, rescue operation, resource allocation, and damage assessment using UAVs during cyclones is described.  Although general commercial UAVs are reported to be used in cyclone operations, UAV systems should be planned specifically for cyclone operations to improve efficiency. Here, the specification required for effective and safe UAV operations in the post-cyclone scenario is presented.  Mission planning required for various rescue, relief, and damage assessment missions related to cyclone management is discussed. A case study of deploying UAV in Amphan cyclone operation in   West Bengal is also presented. This paper can help disaster management authorities to develop UAV systems specifically to cater to cyclone operations. 
\end{abstract}


\section{Introduction}

 % general description 
 
The frequency of natural disasters like cyclones will increase due to climate change, and the government authorities need to prepare a mitigation plan for different stages of disaster. UAVs can be utilized for situational awareness, rescue operation,  communication system, and damage assessment in a disaster scenario.  UAVs pose the advantages of quick deployment, remote accessibility, and the capability of providing high spatial and temporal images and communicating with ground IOTs. Nowadays, quick deployment of UAVs in remote disaster-affected areas is possible due to the rapid advancement of miniaturized UAVs along with the availability of open-source autopilot software such as Ardupilot/ Px4. Application of UAVs in a disaster-related scenario is reported in various literature such as \cite{erdelj2017wireless,panda2019design,malandrino2019planning,kashyap2019uav}. A robust communication platform is required in just the aftermath of the disaster in order to maintain communication with various stakeholders. UAVs can provide communication in case of damage to communication infrastructure. UAV-based IOT platform for disaster management is reported in  \cite{ejaz2019unmanned}. 

In particular, UAVs can be used in the pre-cyclone stage to spread awareness among people in remote locations to provide evacuation instructions on short notice. UAVs can be deployed for survivor detection (\cite{ravichandran2019uav}), weak coast spot detection, coverage mission (\cite{allen2020toward}), and detailed damage assessment (\cite{fernandez2015uav}) for effective resource allocation and rehabilitation plans in the post-cyclone period. UAVs can be equipped for emergency payload transfer to critical areas like hospitals, as access to the area can be blocked for multiple days after a cyclone. Application of UAVs  in cyclone is reported in  hurricane Harvey (\cite{yeom2019hurricane}), hurricane Maria (\cite{schaefer2020low}).

 Damage assessment of cyclones is reported in various literature such as  \cite{haq2012damage, mallick2017living, nandi2020immediate,shamsuzzoha2021damaged, erdelj2017help}.  Damage assessment of Mangrove forests due to Typhoon Yolanda in Calauit island through manual surveying is reported in  \cite{malabrigo2016damage, primavera2016preliminary}. The use of unmanned aerial vehicles (UAVs) for mapping and artificial intelligence (AI) to undertake various types of analysis in the areas of disaster management has made its beginning in post-disaster damage, loss, and needs assessment (PDNA) \citep{calantropio2021deep,yuan2018integration, wu2020analysis}. UAVs are able to generate high-resolution micro-level maps that can show features like buildings, roads, embankments, crop areas, plantations, mangroves, etc., with greater accuracy. AI application is able to create a qualitative and quantitative analysis of the
damage and losses to the features mentioned above \citep{luthi2019applying,syifa2019artificial}.  After hurricane Harvey,  structural damage assessment using DJI Phantom 4 pro is performed after capturing the images at height of 80 m and at an overlap of 80 \% \cite{yeom2019hurricane}.   Damage assessment of roads with the captured images from UAV  using convolution neural networks is discussed in  (\cite{bocanegra2021convolutional}). In this case, CNN is used on the images of roads captured from DJI Matrice 600 Pro after the natural disasters, and   AlexNet is reported to be the best network with an average value of 74.07 \%.    In \cite{kakooei2017fusion},  the fusion of oblique images from  UAV  and vertical images from a satellite is proposed for post-disaster assessment.  Damage assessment using neural network involving captured images from UAV is reported in  \cite{duarte2017towards,xu2018use, vetrivel2018disaster, nex2019towards}.

 %  In this paper 
 This paper describes how the advancement in the UAV and associated technological development could be utilized in the different phases of cyclone management at varying levels of authority.  Although UAV is used in cyclone management; however, UAVs are not designed for cyclone-specific operations. In most cases, an additional sensor like a camera or lidar is fitted with the commercial drones and used in disaster operations.  Commercial drones lack the efficiency to operate quickly in cyclone-affected areas due to wind disturbances. Most importantly, these are not equipped with the advanced algorithms required to perform survivor detection and precise payload delivery. This paper describes the  UAV specifications required to operate safely and effectively in a multi-UAV post-cyclone scenario.  The trade-off of the selection of the type of UAVs for different missions related to cyclones is addressed. The planning required for coverage, search and rescue, and payload delivery related to cyclone operations are presented. The software framework required for multi-UAV operations by multiple authorities in the immediate aftermath of the cyclone is also discussed.   We have presented a case study of deploying UAV locally in Cyclone Amphan. 

  The rest of the paper is described as follows:  Section \ref{application} describes the scope and advantages of UAV in cyclone management. The technical specification required for  UAV system design and multi-UAV system to deploy in cyclone operations is discussed in  Section \ref{preparedness}. Planning required to perform UAV operations for specific cyclone-related missions is presented in \ref{resourceallocation}. Section \ref{damageassessment} describes the UAV planning for damage assessment mission. A case study of the deployment of the cyclone is presented in  \ref{casestudy}.

\section{Applications of UAV in cyclone management}  \label{application}

\begin{figure*}[htb!]
      \centering
      \includegraphics[width=\linewidth, keepaspectratio]{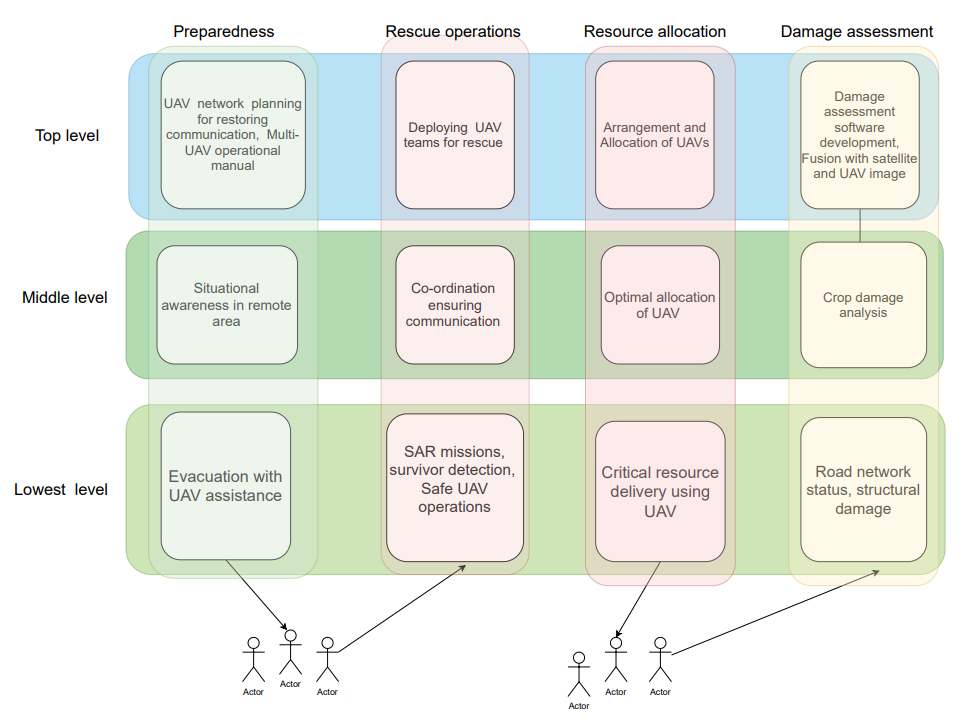}
      \caption{Application of UAV in cyclone management}
      \label{fig:UAV_DSS.png}
   \end{figure*}

There is no standard literature available for the various stages of disaster management. \cite{alexander2002principles}   classified disaster cycle into four phases of prevention, preparedness, response, and recovery; whereas, \cite{erdelj2017help},   classified disaster into three stages:   preparedness,  assessment,  response, and recovery.  In general, disaster management is performed at different hierarchical levels of Government with varying disaster mitigation plans.  We have considered four stages of cyclone management to reflect the varying needs and applications of UAVs at different stages.  The four stages are preparedness, rescue operations, resource allocation, and damage assessment.   Considering the various application of UAVs,  the tasks associated with the UAV  at different stages of cyclone management is presented in Fig. \ref{fig:UAV_DSS.png}. In this case, the three-level hierarchy is considered considering the Indian scenario where the cyclone is managed through a hierarchical level consisting of center, state,  and district authorities. The actors are the people affected in cyclone and might need evacuation/relief assistance. 

At the preparedness stage of the cyclone, the major task involves the prediction of the path of the cyclone, evacuation planning, creating situational awareness, and planning to handle the power and communication interruption.  At this stage, UAVs can be used for situational awareness and evacuation from coastal and remote areas.  UAVs can help create a communication network with the ground IOTs in just the aftermath of a cyclone in case the major communication infrastructure is damaged.  However, the creation of this network will require extensive planning from top-level authorities.  Due to UAVs' easy access and operation, nowadays multiple local authorities independently operate UAVs to get a quick picture immediately after a  disaster. At the preparedness stage, the Government at the topmost level also needs to develop a framework to accommodate the multiple drones by multiple authorities for the safe operation of UAVs in just the aftermath of a cyclone. 
 
 In the rescue operation, important tasks are the deployment of rescue personnel, coordination and maintaining communication among the multiple stakeholders,  and planning search and rescue missions.    At this stage,  UAVs could be used for survivor detection and restoring network connectivity.   Generally, the roads are blocked due to uprooted trees in just the aftermath of the cyclone; therefore, UAVs can be deployed rapidly to access the areas not accessible through traditional means.  Some areas like  island might be cut-off from the mainland, and deployment of a network of UAVs could help restore communication. \\

 In the resource allocation stage, major tasks are the computation of demands, allocation of resources, and distribution of resources. UAVs could be useful to supply relief material to survivors in remote areas.  A  coverage mission by UAVs along the coastal line or covering the cyclone's path might help quickly assess the level of damage and corresponding requirement of relief materials. Critical resources like oxygen and  medicine could be supplied to remote areas using specially designed UAVs.\\

 In the damage assessment stage,  the important tasks are analysis of structural damages, agricultural crop damages,  infrastructure damages, and help in government authorities in the policy-making decision.  Currently,   most damage analysis is performed using satellite images and manual surveying of the affected areas. UAVs fitted with a camera can capture spatial and high-resolution damages of the affected areas at the desired frequency, which could be used for damage analysis using image processing and machine learning techniques.  Images captured from UAVs also could be used to fuse with the satellite images for damage analysis.

\subsection{Advantages of  using UAVs}
 Inclusion of any technology/software in the management of disaster operation should be safe and reliable, flexible for varying levels of operation, quickly deployable on the ground,  ease in handling, and low cost in nature.  The advantages of UAV in disaster management lies in the following aspects.

 \begin{enumerate}
     \item   For any disaster,  the initial few hours are very crucial for survival and rescue operations. UAVs can be deployed in a short time for search and rescue missions. Proper planning of the base station considering the predicted path of the cyclone can immediately deploy the UAV for survivor detection.

     \item    The extent of the damage could be assessed in detail from oblique images captured from UAV which might not be possible to capture from satellite or radar images. Satellite-based damage mapping may not accurately describe the detailed damage of an affected area as fine spatial and high temporal resolution data could be unavailable. For example, the damage due to a surge in water level will happen along the coastal lines so that a UAV survey will capture accurate information rather than satellite image. The images captured from the satellite are costlier and also not readily available.   Also,  sometimes information from satellite data is of limited use because of cloud cover.

     \item   With the rapid advancement in miniaturization of avionics, lightweight structural materials, and open-source drone software, the cost of UAV operations is relatively cheaper than other methods.

     \item  The operation with drones is reliable and does not depend on the existing ground infrastructures that could have been damaged by the cyclone. 
     
     \item   With the advent of various low-cost read-to-fly drones available in the market, drone operation is easy for local authorities with little knowledge of drone technology.

     \item  Mission operational with drones provides flexibility in terms of duration, frequency, time, and scale of operation.

     \item Operation with UAVs ensures the accessibility to remote areas. 
     
 \end{enumerate}

\subsection{Selection of UAV}
   During pre and post-cyclone operations, the primary tasks are coverage, search and rescue,  and payload delivery. In the case of a cyclone,  both the fixed-wing and rotary-wing should be deployed based on the mission requirement.  A coverage mission will require a long-endurance UAV, whereas a critical payload delivery mission requires hovering capability.  Apart from the specific mission requirement, 
   the selection of the type of UAV needs to be performed considering payload carrying capability, the requirement of the landing area,  and operational cost.   Fixed-wing UAVs will have advantages in terms of long endurance, high-speed coverage, aerodynamic efficiency; whereas, multi-rotor will have advantages of precise payload delivery and lesser take-off/landing area requirement. The technical knowledge required to operate a fixed-wing is higher than a multi-rotor. Considering the obstacles of uprooted trees, damaged power, and communication lines, the operation of a fixed-wing is more difficult compared to a multi-rotor at a lower height.   The main bottleneck in deployability of fixed-wing in cyclone-affected areas could be of non-availability of take-off and landing area, whereas multi-rotor could be deployed from almost from any location.   Hand launch fixed-wing UAVs could be deployed; however, this required skilled personnel to launch and catch the UAV during landing.

   Coverage missions will require long endurance and high-speed UAV for quick coverage of the damaged area. Search and rescue missions also require long endurance and high-speed UAVs; however, SAR missions need to be planned at a lower height, and it requires hovering around the survivor to gather detailed information.   Payload delivery mission of general items requires a UAV with high payload capability; however, delivering a critical item to a survivor/hospital might require precise delivery. Situational awareness/ evacuation planning with UAV  might require the hovering capability of UAV along with high speed.    In the case of structural damage assessment, UAVs need to operate at low altitudes in cluttered environments with uncertain obstacles; however, agriculture damage assessment could be performed at higher altitudes with high speed to cater to larger areas.   Considering all the above facts, it is recommended that coverage mission and general relief mission could be planned using fixed-wing UAVs,  whereas SAR mission and critical payload delivery mission with multi-rotor UAV.  However, an initial coarse SAR mission could be planned with fixed-wing UAV quickly, and after assessment of initial damage,  a finer SAR mission using multi-rotor UAV could be planned for specific areas.
   In the case of evacuation missions and damage assessment missions, the selection should be made considering the specific evacuation requirement.   Therefore, a UAV fleet deployed in cyclones should consist of both the fixed-wing and multi-rotor wings to optimally cater to the requirement of different missions. Specifically, a UAV fleet with only a fixed-wing should not be considered to avoid a long time of deployment in case of extensive damage to the take-off area.

\section{Specification of UAV systems} \label{UAVspec} 
UAVs with desired hardware and software capability and detailed mission planning are required for efficient cyclone management.  UAVs need to be designed/procured considering the specific operational requirement of autopilot functionalities, payload-carrying capability, and sensor requirement for cyclone management.     The important  basic technical specification of UAVs for cyclone operation   are listed as follows:  
 
  \begin{enumerate}
  
    \item   UAVs should be able to capture images/videos while performing autonomous waypoint missions. It should be equipped with a minimal sensor suite of  IMU, altimeter, GPS, and camera. 
    
    \item  UAVs used for precise payload delivery and survivor detection should have the capability of processing images on board in real-time. 
    
      \item  Generally, the wind speed is high a few days before and after the cyclone; UAVs should withstand high wind and gust to be operational during the critical evacuation and rescue operation period.

      \item UAVs should be equipped with the Inter-agent collision avoidance algorithm and obstacle avoidance algorithm for safe operations in a complex multi-drone mission by multiple authorities. 
         
       \item  The UAV should have an algorithm to restrict inside the Geofence area to remain in the designated corridor without violating the statutory regulations. 
       
  \end{enumerate}

Apart from the basic specifications, UAVs could be equipped with thermal imaging sensors for the detection of survivors. The capability to transmit the imagery to the ground station in real-time would be better for survivor detection. The UAVs used for a longer coverage mission should have sufficient capacity of storage for the images/ videos. The efficient operation of multiple UAVs requires coordination among UAVs, optimal task allocation to individual UAVs, and collision avoidance. Multiple UAVs should be operated in a centralized architecture for optimal use; however, communication with the base station/leader UAV should be kept at a minimal. The multi-UAV software architecture should have the following specifications for safe and efficient operations. 

   \begin{enumerate}
       \item  It should be scalable to handle missions with few UAVs to missions with a large number of UAVs. 
       
       \item  It should be flexible to accommodate the different types of UAVs. 
       
   \end{enumerate}

Additionally,  some of the UAV fleets should also be capable of operating at night. UAVs should be deployed with the area's topology in mind, as some UAVs may struggle to perform well in high-altitude environments.

\section{Preparedness stage}  \label{preparedness}
  At the preparedness stage, apart from situational awareness and evacuation planning using UAV,  the UAV specific tasks are UAV system design and development of regulatory framework required for UAV operations during the disaster. 

\subsection{UAV system design}
 UAVs should be designed keeping in mind the specification listed in Section \ref{UAVspec}.  The UAV can be designed in a specific configuration such as hexacopter to improve wind resistance capability. A robust and adaptive controller  should be used to handle the high disturbances from wind gusts (\cite{fernandez20171,jana2017composite}).  The main autopilot, along with the vision processing module, should have the capability of human detection,  survivor tracking, and payload delivery. 

 In case of a UAV fleet designed with commercial UAVs, the vehicle needs to be reconfigured for the advanced mission as commercial UAVs are mostly equipped only for autonomous waypoint mission.  These types of UAVs need to be equipped with a vision processing module for onboard image processing required for SAR and payload delivery missions.  The vision processing module should have the computation capability to process the image in real-time for SAR missions. The mission algorithm should be developed considering the compatibility with the existing code of commercial autopilot. The software architecture should be designed to incorporate the commercial UAVs with little modification. Software architecture proposed by \cite{agnel2021autonomous}  for multi-drone mission could be used for integration of autopilot of  commercial UAV.  This will also help in scaling the UAV operations with  different UAV systems.

 \subsection{UAV operations}
 % Put into cyclone scetion 
          The disaster control room should have a  dedicated unit to handle the complex operation of multiple drones by multiple authorities. It should clearly mention the geo-fence restrictions for each drone and local station to avoid collision among the drones. If the UAV operation is planned beforehand, then the  UAV decision support system (DSS)   framework should be able to deploy the drones in the overall resource and task allocation framework. DSS framework for resource and task allocation proposed by \cite{jana2021decision} could be used in this purpose.  The overall software framework should be able to assign the drones dynamically in the desired area. The DSS should include GUI  to monitor the location of drones and analysis of live footage. 
    
          It might not be possible to plan UAV operations in advance using a central base station in many cases. In many cases,  local pilots might be flying manually to assess the ground situation.    Flying drones by multiple people can create chaos and also risk the chance of collision avoidance. So, a  software framework is required for coordination among multiple pilots, providing permission from the government authority and addressing the local people's concerns due to UAV operations.  The software framework should include the pilot information and desired flight plan for the permission of the disaster management  authority.

 \section{Rescue operations and resource allocation}    \label{resourceallocation}        
 In the rescue operation stage, deployment of UAV  needs to be planned for coverage missions, SAR missions, payload delivery, and damage assessment. The planning related to UAV design, and algorithm requirements are discussed in this section.   The overall UAV operations can be handled in different hierarchical operation layers through the main base station and multiple local base stations.

\subsection{Coverage mission}
      Total UAVs should be distributed into the different base stations instead of a single base station for better coverage.  The expected disaster-affected areas should be classified into different zones based on the expected spread of the cyclone affected areas,  available UAVs, the endurance of UAVs, topology of the area, and various other local factors such as logistics road network, availability of the suitable site.  A dynamic partitioning algorithm could also be considered to balance the load on UAVs with continuous assessment of each area's coverage requirement. Each base station should be assigned a particular area.  The area under a base station should be further divided based on the number of  UAVs for coverage missions under that station.  Voronoi partitioning algorithm based on endurance could be considered for allocation.  Overall mission planning for each UAV to be designed considering endurance of UAV, speed of UAV and desired overlap in captured images.  The area under each base station is to be allocated to UAVs through optimal task allocation, and the corresponding waypoints are to be loaded in the autopilot of each UAV. The images captured from the UAV could be stored or transmitted to the base station.

  \subsection{SAR mission}
     Application of UAV in search and rescue operation after the disaster could help in rapid detection of survivors and deployment of relief in a short time. The search and rescue mission will be planned as a coverage mission; however, a UAV will be assigned once an object of interest is detected. Onboard vision processing module to be equipped with human detection or any specific object detection. Machine Learning-based algorithms could be used for the detection of humans after training with available images from a similar disaster location.

\subsection{ Relief distribution}
      Generally, post-cyclone relief distribution involves providing dry food packets and medicines to an area; however, in some cases, payload needs to be dropped precisely near a survivor standing on a boat, roof, or a critical medical item like blood/ oxygen to a  hospital in a remote area.  General payload delivery could be arranged through fixed-wing and multi-rotor to be used for precise delivery. For delivery using a fixed-wing, it should be fitted with a mechanism to trigger the release of payload based on the GPS location.  The error in GPS location could be around 1-2 m, and payload to be planned to be dropped over an area.   In case of precise delivery with a multi-rotor wing, it should have the capability to detect the specific target (survivor/landing pad) and then release the payload or land at the desired location. The UAVs are to be designed to carry the desired payload and the controller to be designed considering the possible variation of mass and inertia over the flight regime. Apart from direct payload delivery, the mission needs to be designed over the pre-cyclone road network to assess the current road status for relief delivery through conventional transport.

\section{Damage assessment}  \label{damageassessment}
    
    In the case of a cyclone, damage assessment include agricultural damage,  structural damage, mangroves damage.  
   The damage assessment mission needs to be planned at different layers from coarser to finer coverage based on the assessment after each layer.   The altitude of the operations should also be decided based on the camera specifications and terrain characteristics. Damage assessment is mostly performed from the captured images from the camera; however, other sensors such as lidar/radar could be used for specific requirements.  Locally damage analysis could be performed using  visual cues after generating the orthomosaic images; however, for a large region, an automated software integrated with machine learning techniques has to be used for quick assessment.  For advanced analysis, the mission needs to be planned carefully considering the illumination, the incident angle of the camera, the field of view of the camera, and desired overlap. Cameras need to be calibrated properly as the analysis results are sensitive to the camera calibration parameters.    Generally, a mission is to be planned as a straight and level flight; however, it will be difficult to maintain throughout the flight duration due to disturbances such as wind gusts. So, the vehicle states parameters also need to be stored at the desired frequency to obtain the exact location and orientation of the captured images.  Vehicle state parameters include its position, velocity, Euler angles, and angular rates.  Currently, most commercial UAVs allow to store these parameters at a  ground station; however, the flight logs are sometimes unavailable due to communication failure.   To obtain high accuracy in damage analysis,  the flight parameters need to be stored onboard at the desired frequency. The captured images need to be geo-tagged for better analysis.   Although a few images from disaster locations are available,  authorities might need to create training databases after ground survey aftermath of a disaster for future applications.

 \section{Case study} \label{casestudy}
 In order to implement the large-scale deployment of UAVs in cyclone situations,  we have deployed UAVs locally during cyclone Phani (2019), Bulbul(2019), and Amphan (2020) to understand the ground complexities involved in UAV operation during the cyclone.  We performed preliminary damage analysis from the captured images and augmented the training dataset to be used for future cyclones. 
 
\begin{figure}[ht!] 
       \centering
       \includegraphics[width=0.7\columnwidth, keepaspectratio]{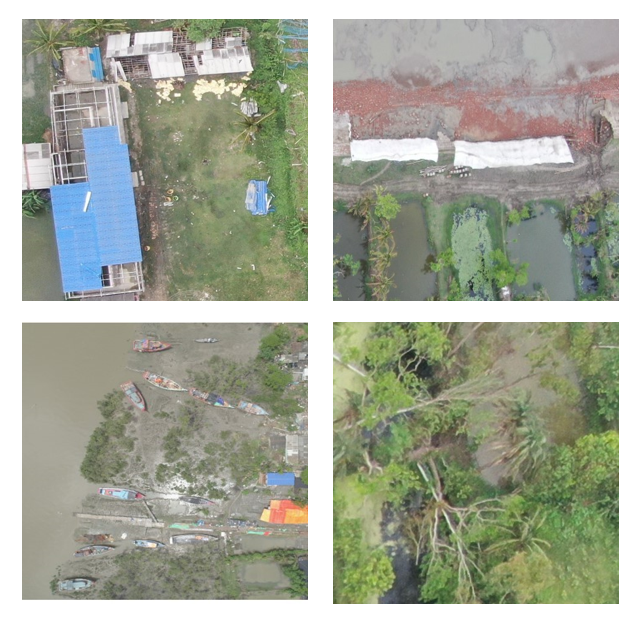}
       \caption{Damages captured by UAVs in West Bengal: damaged building, damaged Jetty, uprooted trees,  damaged embankment (anti-clockwise from top corner) }
       \label{damaged}
\end{figure}

 Cyclone Amphan, the most powerful storm in the Bay of Bengal in over a decade, ripped through West Bengal (Kolkata, North 24 Parganas, South 24 Parganas Hooghly, Howrah, and East Medinipur) as well as  Bangladesh. Amphan's landfall process commenced from the afternoon (20th May 2020)  to early morning (21st May 2020), causing enormous damage to people's lives, infrastructure, agriculture, and property. We have deployed drones in 30 locations covering 11-gram panchayats of district South-24 Pargana in state WestBengal of India. The  UAV operation is performed at the height of 80 m. The endurance of the vehicle was around 25-30 minutes.  UAV is fitted with 4k HD vision camera.  Sample images captured by the UAV is shown in Fig. \ref{damaged}. Visual analysis of these raw images could provide a  quick assessment of the urgent relief/rescue operations.  

\begin{figure}[h!]
       \centering
       \includegraphics[width=0.6\columnwidth, keepaspectratio]{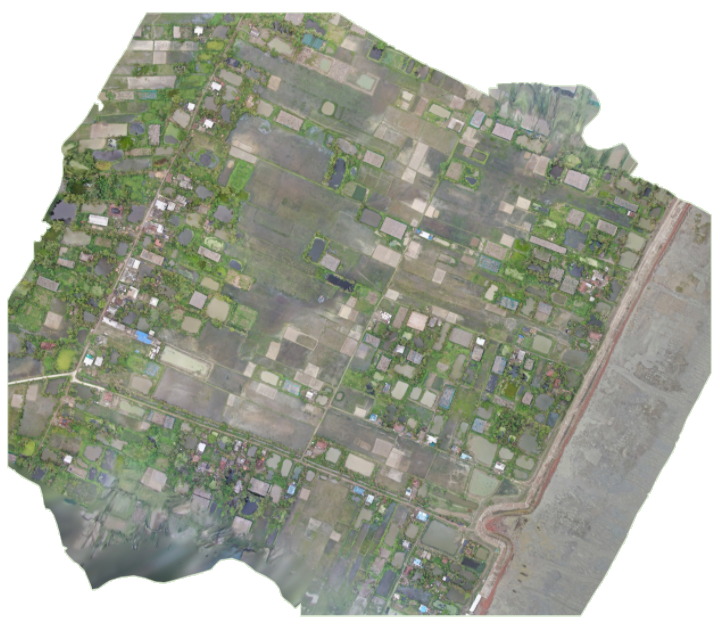}
       \caption{Sagar Island, West Bengal}
       \label{ortho}
\end{figure}

\begin{figure}[ht!]
       \centering
       \includegraphics[width= 1.0\columnwidth, keepaspectratio]{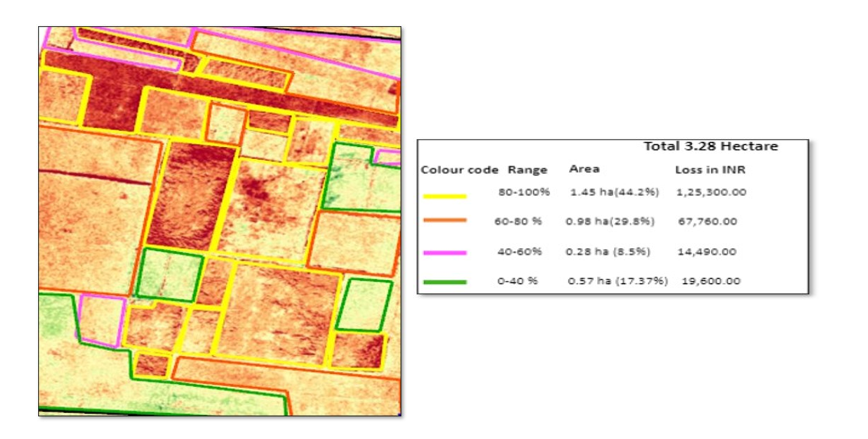}
       \caption{Analysis of crop damage at Ganga Sagar in West Bengal}
       \label{crop_d}
\end{figure}

The images are further processed in Invent Grid's photogrammetry engine. Images are further stitched to obtain an orthomosaic view to obtain a better idea of a large area. Fig. \ref{ortho}  shows an orthomosaic view of Sagar island after stitching the images captured from the UAV. A  preliminary report on crop damage analysis is shown in Fig. \ref{crop_d}.  The Colour code shows the boundary colors of the patches of the agricultural field. The following lesson is learned for future deployment of UAVs in cyclone operation. 
\begin{enumerate}
    \item  Sufficient datasets of cyclone-affected images are required to improve the accuracy of ML algorithms. 
    
    \item  Wind is the crucial factor to be taken into the design.   Currently, we could operate only at some specific duration of the day. 
    
    \item   A great detail of logistical planning (Battery Charging, base station, etc.)  is required to deploy drones in remote cyclone-affected areas as existing infrastructure is damaged. 
    
    \item  Privacy issues and government regulations for deploying drones in cyclone-affected civilian areas need to be addressed. 
    
    \item Collaboration for humanitarian actors and experts from drone operations need to be enhanced for better efficiency. 
\end{enumerate}

 \FloatBarrier
 
\section{Conclusions}
 This paper presents a planning framework for the application of UAVs in cyclone management. Details of planning related to the UAV design stage to the UAV  deployment stage at various stages of cyclone operation are addressed.   Different hardware and software specifications of UAV systems for deployment in cyclone management are listed. Cyclone-specific coverage mission, search and rescue mission, damage assessment mission is presented. Initial experiences of deployment of UAV in cyclone situation shows that UAV can be utilized to accurately map the cyclone-affected areas. 
\FloatBarrier

\section*{Acknowledgement}          
The researchers from the Indian Institute of Science and University of Exeter acknowledge partial funding support from the EPSRC-GCRF project  "Emergency flood planning and management using unmanned aerial systems" (EP/P20839X/1).

Humanitarian Aid International (HAI) acknowledges the HCL Foundation, India for funding  and Sabuj Sangh for logistics support during the project work. The data collection using drones has been carried out in partnership with Invent Grid Pvt. Ltd., based in Bhubaneswar, India, under the direction of HAI. Invent Grid has also provided their expertise in data processing, output generation, DEM, and presentation of the outcome.

\bibliographystyle{apalike}
\bibliography{main}

\end{document}